\documentclass[twocolumn,appendixfloats,tighten]{aastex6}
\usepackage{newtxtext,newtxmath}
\usepackage[T1]{fontenc}
\usepackage{ae,aecompl}

\usepackage{amsmath}	
\usepackage{amssymb}	

\usepackage{ctable}
\usepackage{url}
\usepackage{xspace}
\usepackage[normalem]{ulem}
\usepackage{hyperref}
\usepackage[all]{hypcap}
\usepackage{graphicx}

\newcommand{\be}{\begin{equation}}
\newcommand{\ee}{\end{equation}}

\newcommand{\tmin}{t_{\rm min}}
\def\h2{${\rm\,H_2}$}

\usepackage{color}

\begin{document}

\title{Measuring the delay time distribution of binary neutron stars. III.  Using the individual star formation histories of gravitational wave event host galaxies in the local universe}  
\author{Mohammadtaher Safarzadeh\altaffilmark{1,2}, Edo Berger\altaffilmark{1}, Joel Leja\altaffilmark{1} and Joshua S. Speagle\altaffilmark{1}}

\altaffiltext{1}{Center for Astrophysics | Harvard \& Smithsonian, 60 Garden Street, Cambridge, MA, 02138, USA,
  \href{mailto:msafarzadeh@cfa.harvard.edu}{msafarzadeh@cfa.harvard.edu}}
\altaffiltext{2}{School of Earth and Space Exploration, Arizona State  University, AZ, USA}

\begin{abstract}
In Paper I (Safarzadeh \& Berger 2019) we studied the determination of the delay time distribution (DTD) of binary neutron stars (BNS) through scaling relations between halo/stellar mass and the star formation history (SFH) of galaxies hosting gravitational wave (GW) events in the local universe.  Here we explore how a detailed reconstruction of the individual SFHs of BNS merger host galaxies can improve on the use of the scaling relations. We use galaxies from the Galaxy and Mass Assembly (GAMA) survey, which is mass complete at $M_*>10^9$ M$_\odot$ in the redshift range $0.05<z<0.08$. We use the reconstructed SFHs derived from the {\tt Prospector} code, for two distinct sets of priors (favoring continuous and bursty SFHs), and convolve those with power law DTDs characterized by an index $\Gamma$ and a minimum delay time $\tmin$.  We find that with this approach $\mathcal{O}(100)-\mathcal{O}(300)$ host galaxies are required to constrain the DTD parameters, with the number depending on the choice of SFH prior and on the parameters of the true DTD.  We further show that using only the host galaxies of BNS mergers, as opposed to the full population of potential host galaxies in the relevant cosmic volume, leads to a minor bias in the recovered DTD parameters.  The required host galaxy sample size is nearly an order of magnitude smaller relative to the approach of using scaling relations, and we expect such a host galaxy sample to be collected within a decade or two, prior to the advent of third-generation GW detectors.
\end{abstract}

\section{Introduction}

The delay time distribution (DTD) of binary neutron stars (BNS) is currently poorly constrained, but as we have recently shown it can be determined using both the mass distribution of BNS merger host galaxies in the local universe (\citealt{sb19}; hereafter, Paper I) and the redshift distribution of BNS mergers as probed by third-generation gravitational wave (GW) detectors (\citealt{sb+19}; hereafter, Paper II). The former approach takes advantage of galaxy scaling relations that map halo/stellar mass into star formation history (SFH), which when convolved with the DTD lead to a predicted BNS merger host galaxy mass function (this approach was previously proposed and used in the context of short GRBs: \citealt{Zheng:2007hl,Kelley2010,lb10,fbc+13,Behroozi:2014bp}). The host galaxies of BNS mergers can be identified through the detection of electromagnetic (EM) counterparts, but this is likely only achievable within a few hundred Mpc.  We found that for a power law DTD characterized by index $\Gamma$ and minimum delay $\tmin$, $\mathcal{O}(10^3)$ host galaxies are required to reasonably constrain the DTD.

The latter approach instead relies on a redshift mapping of the BNS merger rate, which requires GW detections to $z\sim{\rm few}$, achievable with the next-generation Einstein Telescope (ET) and Cosmic Explorer (CE).  In this approach, it is unlikely that EM counterparts can be detected, but the individual redshift uncertainties from the GW data ($\delta z/z\approx 0.1z$) can be overcome through a large number of anticipated detections, $\sim 10^5$ yr$^{-1}$.  We found that with about a year of CE+ET data the DTD parameters, as well as the mass efficiency of BNS production, can be determined to about 10\%. 

Here we continue our investigation of the DTD, with an alternative approach to the use of BNS merger host galaxies at $z\approx 0$.  Namely, unlike in Paper I, which used scaling relations between mass and SFH, we explore the use of detailed reconstructed SFHs for the individual host galaxies.  In \S\ref{sec:gama} we present the galaxy sample used for this study (the Galaxy and Mass Assembly survey) and its SFH reconstruction. In \S\ref{sec:method} we present the method for extracting the DTD from the galaxy SFHs, as well as our approach to evaluating the number of host galaxies required.  In \S\ref{sec:results} we discuss our findings in terms of the sample size needed, and we conclude in \S\ref{sec:conc}.  We adopt the Planck 2015 cosmological parameters \citep{Collaboration:2016bk}: $\Omega_M=0.308$, $\Omega_\Lambda=0.692$, $\Omega_b=0.048$, and $H_0=67.8$ km s$^{-1}$ Mpc$^{-1}$.

\section{Galaxy Data}
\label{sec:gama}

\begin{figure*}
\centering
\includegraphics[width=0.32\linewidth]{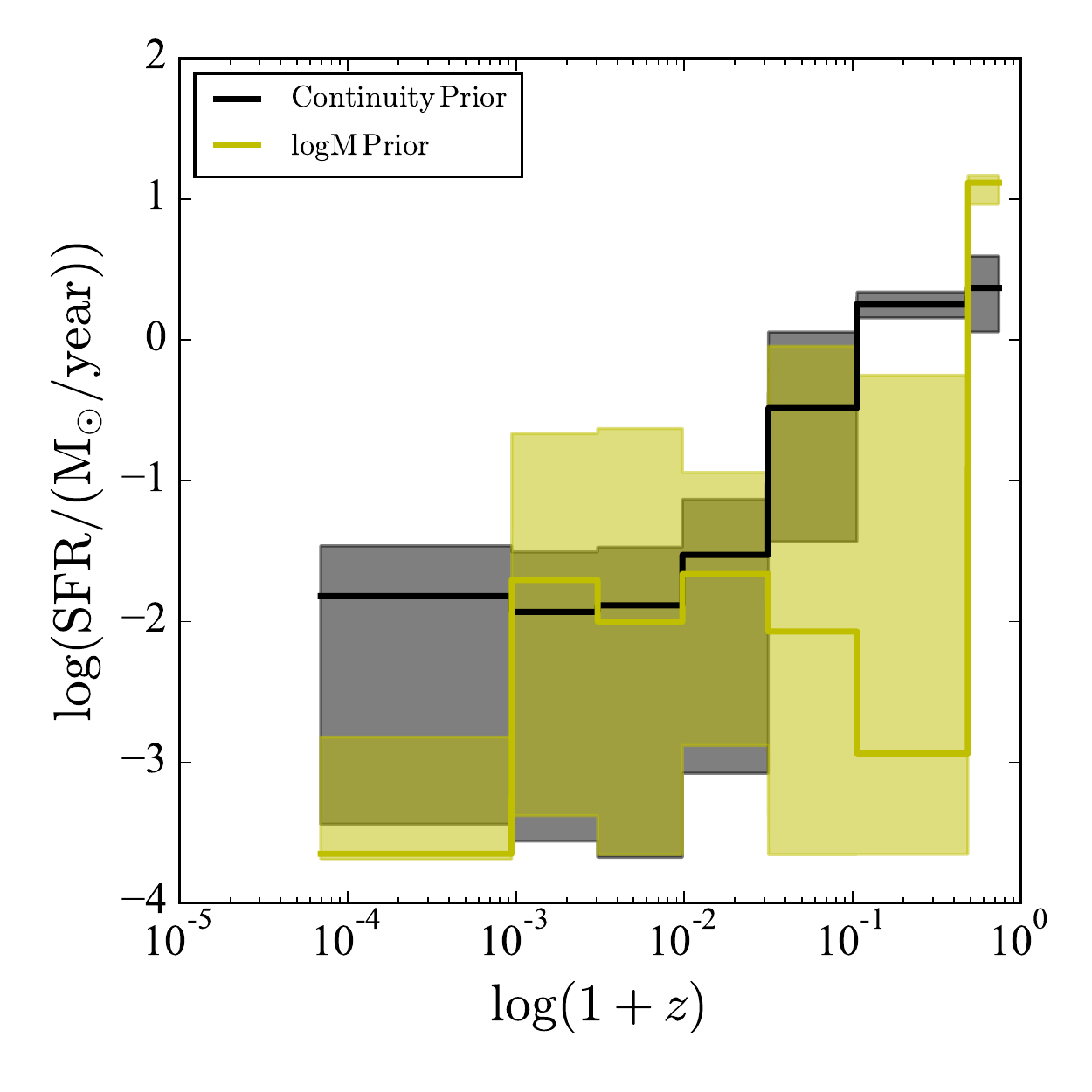}
\includegraphics[width=0.32\linewidth]{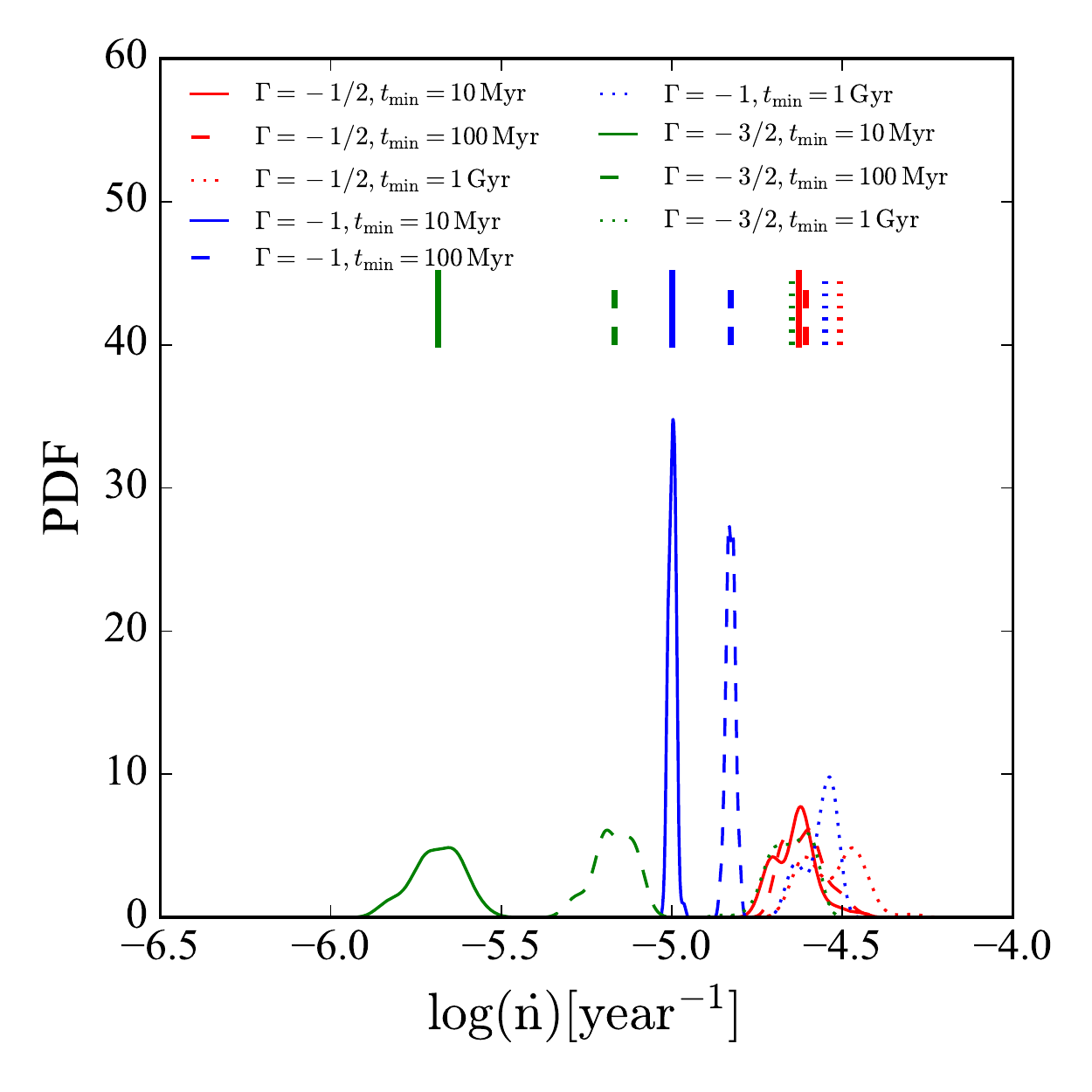}
\includegraphics[width=0.32\linewidth]{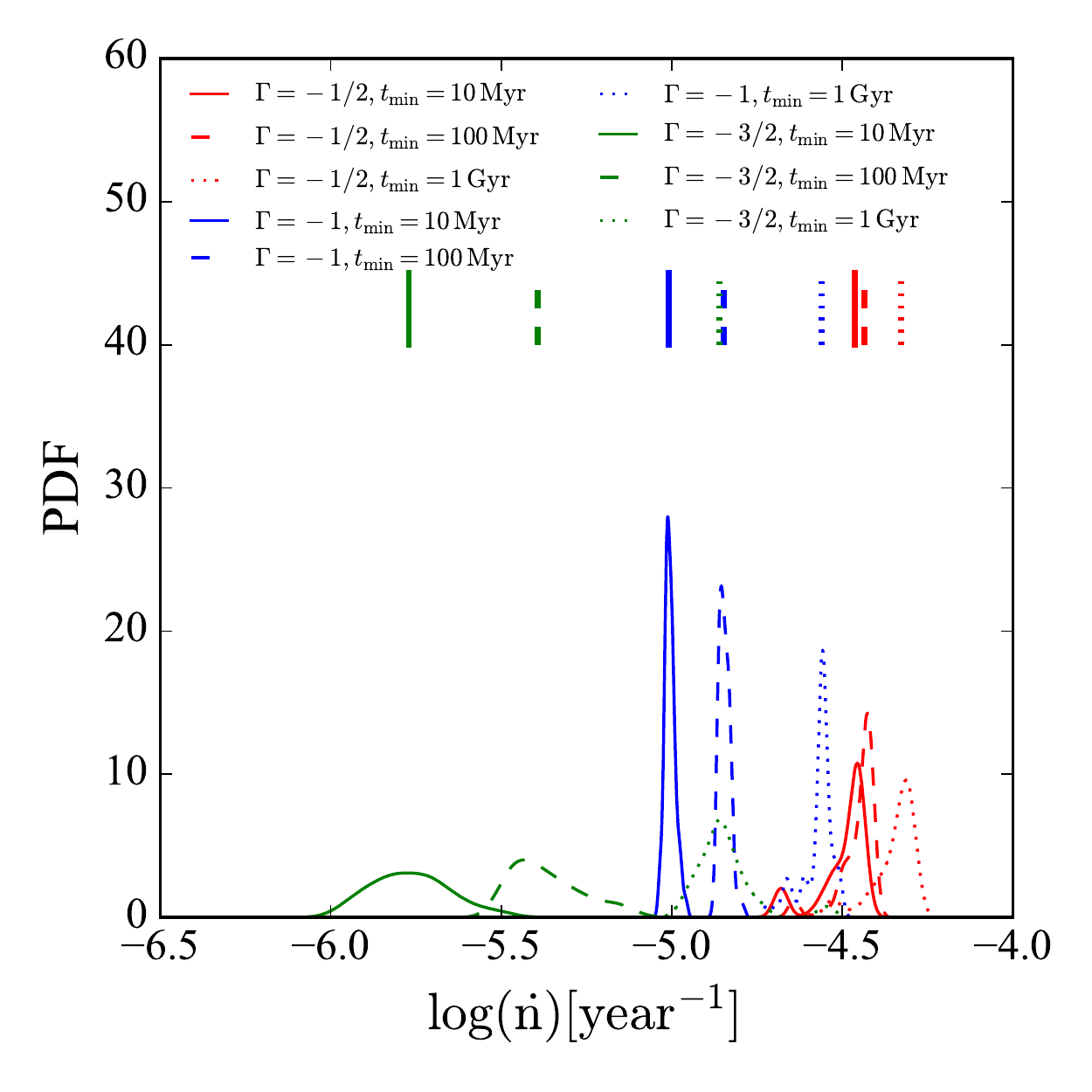}
\includegraphics[width=0.32\linewidth]{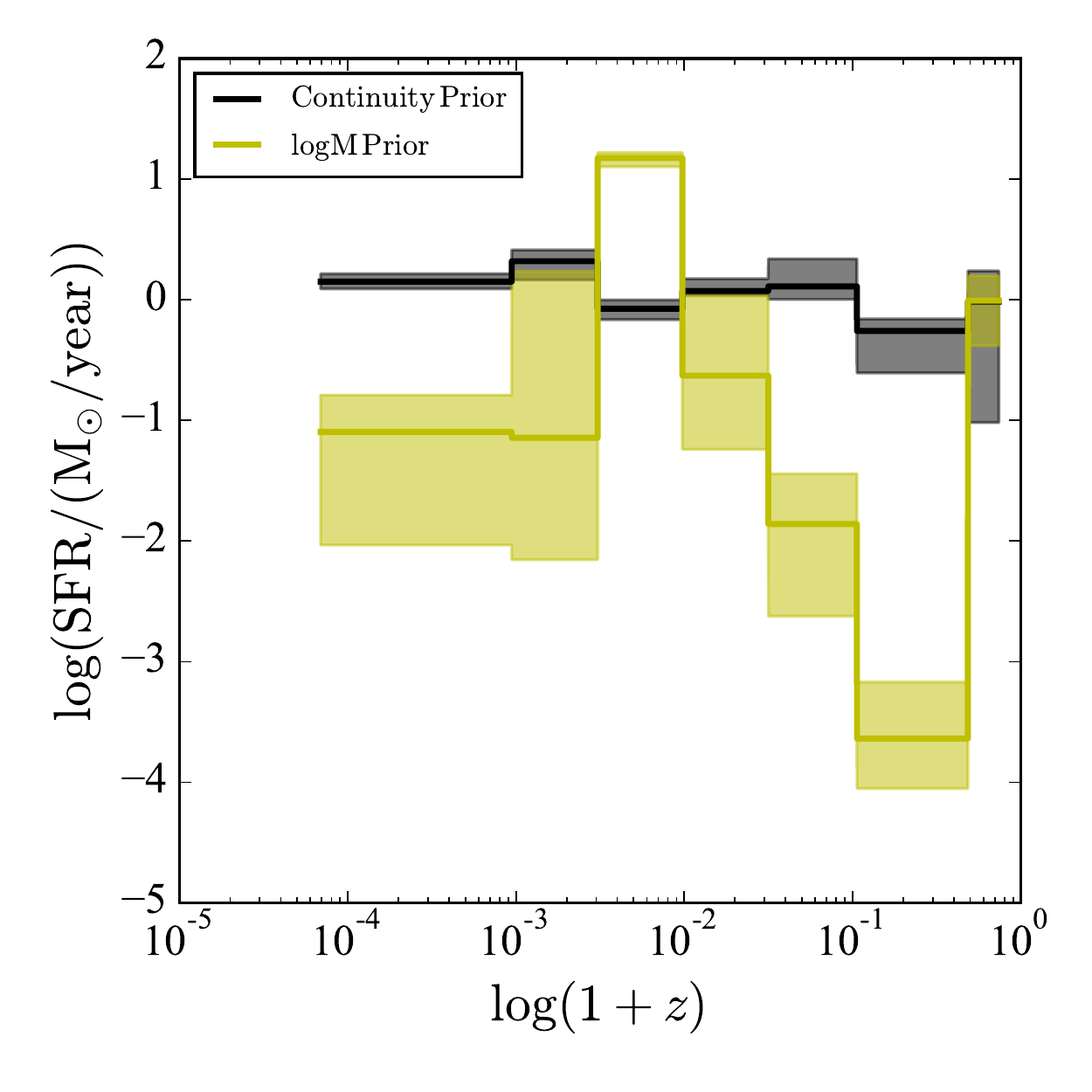}
\includegraphics[width=0.32\linewidth]{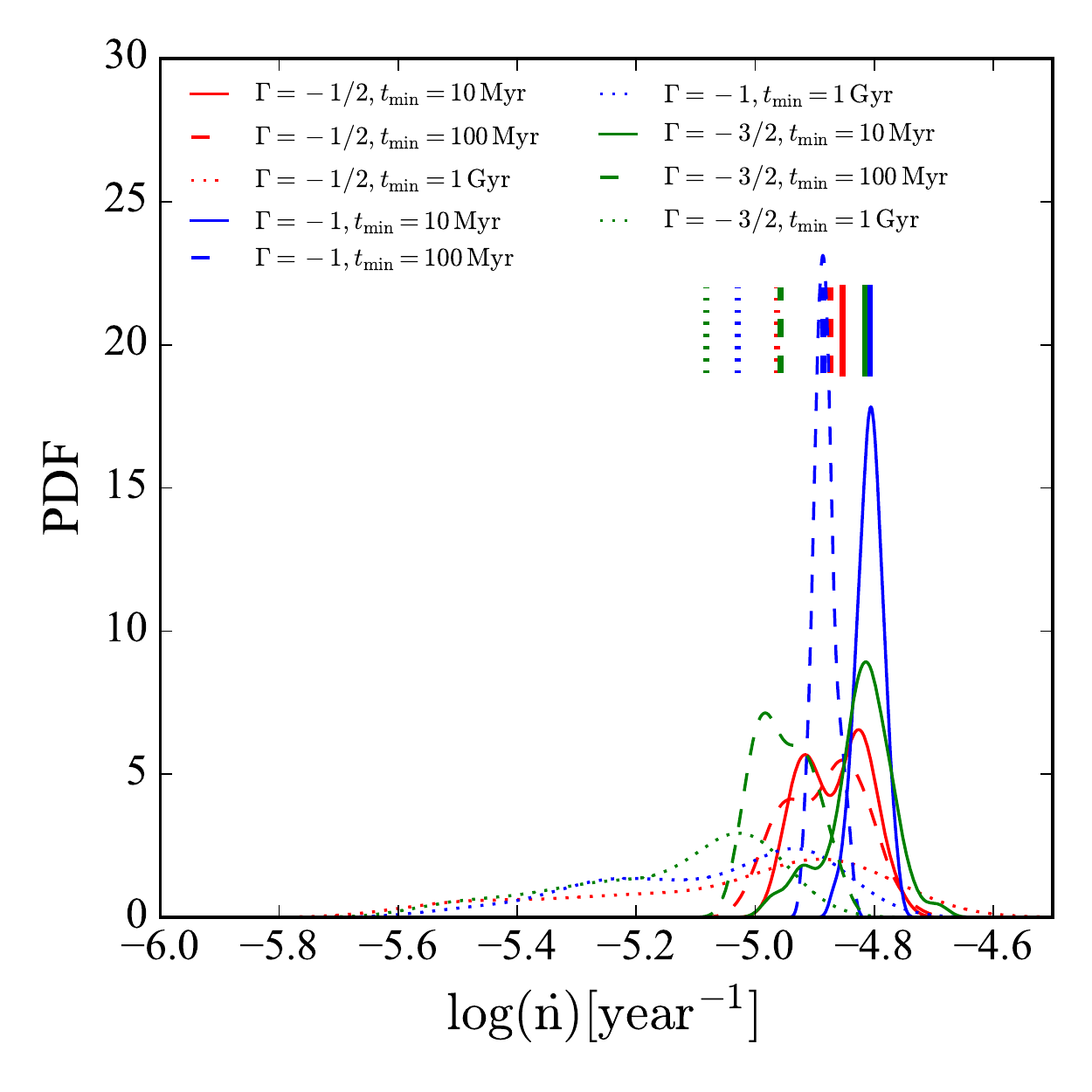}
\includegraphics[width=0.32\linewidth]{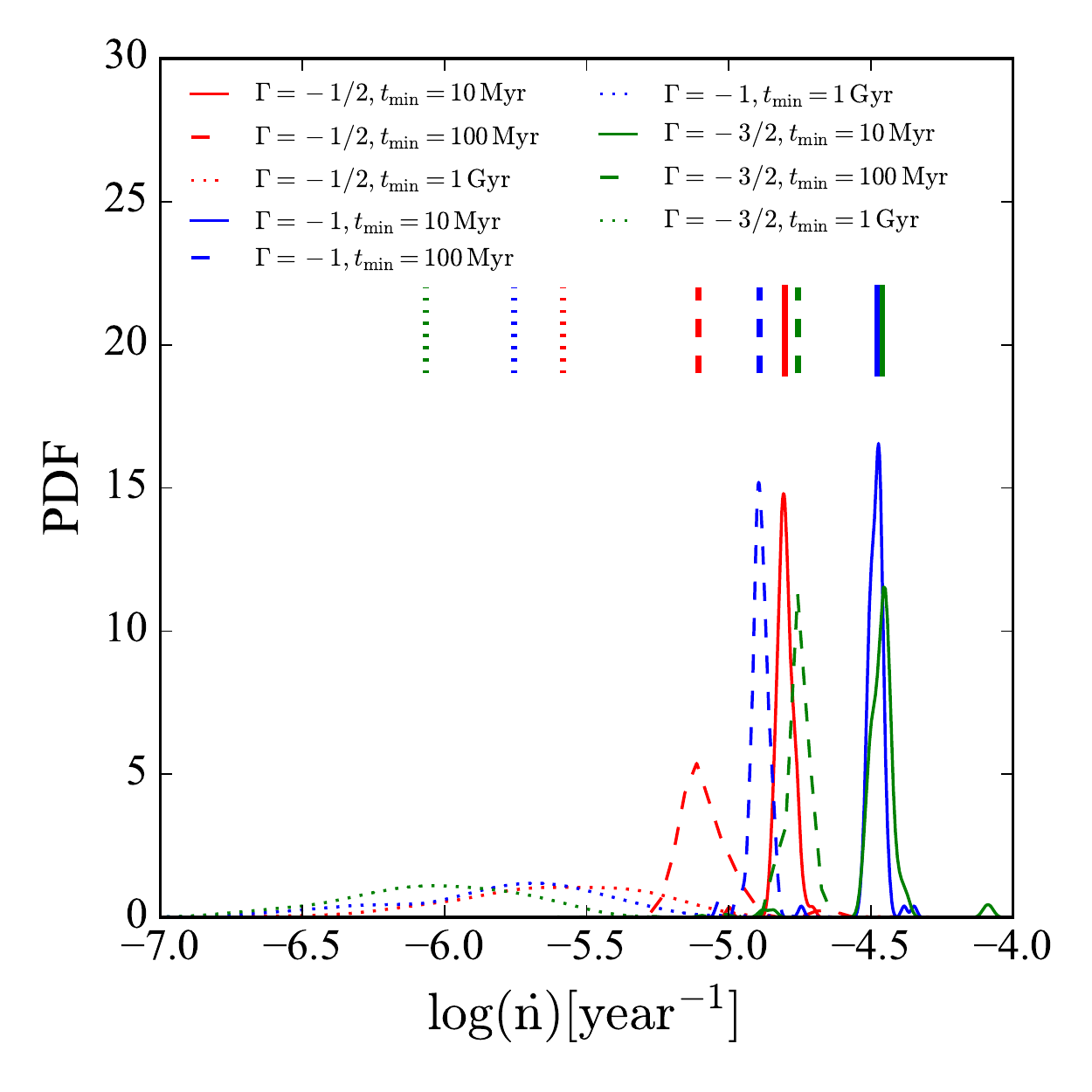}
\caption{The BNS merger rate for two galaxies from the GAMA survey with distinct SFHs ({\it Left}) that peak at early ({\it Top}) and later ({\it Bottom}) cosmic time (based on Equation~\ref{eqn:ndot}).  We show the SFH based on the two types of priors (yellow: logM; grey: continuity), with the solid line indicating the median SFH and the shaded regions marking the range of 16th to 84th percentile. {\it Middle} and {\it Right} are the merger rate PDFs from convolution of the 9 DTDs with the posterior distribution of SFH of the galaxy modeled based on continuity and LogM priors respectively. The vertical bars indicate the median values of the merger rate PDFs for each DTD, shown with the same line style and color.}
\label{f:priors}
\end{figure*}

We use SFHs inferred from galaxy photometry in \citet{leja19}. The photometry is measured with the LAMBDAR code \citep{wright16} from DR3 of the Galaxy and Mass Assembly (GAMA) survey \citep{driver11,baldry18}, and includes 21 bands ranging from the far-UV ({\it GALEX}) to the far-IR ({\it Herschel}). The galaxies have spectroscopic redshifts. All galaxies at $0.05<z<0.08$ with stellar masses $M_*>10^9$ M$_{\odot}$ as determined in \citet{taylor11} are modeled, resulting in a mass-complete sample of 6134 galaxies.

The photometry was fit using the Prospector-$\alpha$ model within the \texttt{Prospector} inference machine \citep{leja17,prospector17}. This model includes a seven-parameter non-parametric star formation history, as well as flexible dust attenuation model, far-infrared dust re-emission via energy balance, and nebular emission self-consistently powered by the stellar ionizing continuum. 

A key source of uncertainty in SFH recovery is the choice of prior \citep{carnall19,leja19}. This sensitivity occurs because the SEDs of stellar populations change slowly as a function of time; specifically, SEDs evolve roughly evenly in each logarithmic time step \citep{ocvirk06}. As a result, the SFR inferred in adjacent time bins is typically highly degenerate. Fortunately, key outputs of galaxy SED-fitting such as the mass-to-light ratio and, to a lesser extent, the recent SFR, are calculated using moments of the SFH, which are largely insensitive to this degeneracy \citep{bell03,leja19}. However, the DTD does not interact with these conserved quantities but instead couples directly to the SFH, and is thus sensitive to this degeneracy. Accordingly, we perform the analysis here using two different priors that assume opposite behaviors in this degeneracy: one that favors bursty SFHs (hereafter, logM) and one that favors smooth SFHs (hereafter, continuity) \citep{leja19}.  Two representative examples for galaxies with opposite SFHs are shown in Figure~\ref{f:priors}, for both the logM and continuity priors.  These examples highlight both the range of behavior and the associated uncertainties (statistical and systematic) in the reconstructed SFH.

The two SFH priors adopted in this work encapsulate the plausible range of choices \citep{leja19}. 
Star formation histories derived from high-resolution optical/near-IR spectroscopy can provide more precise SFHs than photometry alone, 
although in most cases this will be a modest improvement \citep{2012MNRAS.421.2002P}.

\section{DTD Determination Method}
\label{sec:method}

The expected BNS merger rate at $z=0$ for galaxy $i$ with star formation history $\psi_i(z)$ is given by:
\begin{align}
  \dot{n}_i=&\int_{z_b=10}^{z_b=0}
  \lambda\frac{dP_m}{dt}(t-t_b-t_{\rm
    min})\psi_i(z_b)\frac{dt}{dz}(z_b)dz_b,
    \label{eqn:ndot}
\end{align}
where $dt/dz = -[(1+z) E(z) H_0]^{-1}$ and $E(z)=\sqrt{{\Omega}_{m,0}(1+z)^3+{\Omega}_{k,0}(1+z)^2+{\Omega}_{\Lambda}(z)}$; $t_b$ is the cosmic time corresponding to $z_b$; $\lambda$ is the BNS production mass efficiency, assumed to be a fixed value of $10^{-5}$ M$_{\odot}^{-1}$, independent of redshift or environment; and $dP_m/dt$ is the DTD, likewise assumed to be independent of environment.  
As in Papers I and II, we parameterize the DTD to follow a power law\footnote{In this work (as well as in Papers I and II) we have focused on a power law DTD, which is well motivated.  
However, it is possible that the true DTD might deviate from this power law form. 
This could potentially be investigated by comparing the results of the analysis proposed here 
(using host galaxies at $z\sim 0$) and the analysis in Paper II (using the redshift distribution of BNS mergers from third-generation detectors).} 
with index $\Gamma$, minimum delay time $t_{\rm min}$, and a fixed maximum delay time $t_{\rm max}=10$ Gyr. Our results are not sensitive for a larger value of $t_{\rm max}$. This formulation of Equation~\ref{eqn:ndot} is identical to that used in Paper I, except that there we took $\psi$ to be a direct function of halo mass, $M_h$.

To generate the simulated data, we assume a fixed number of $N_{\rm gal} = 1000$ galaxies that can serve as possible hosts for BNS merger events. These are selected to serve as a representative subset of the GAMA sample that preserves the mass distribution of the full sample. For a given DTD ($\Gamma$ and $t_{\rm min}$) and SFH ($\psi_i(z)$) we can then estimate the mean merger rate, $\dot{n}_i$, for each galaxy using Equation~\ref{eqn:ndot}. This then gives each galaxy a different probability to host a BNS merger event for the different DTDs (see Figure~\ref{f:priors}).  As in Papers I and II, we use a set of 9 representative DTDs, with $\Gamma=[-1.5,-1,-0.5]$ and $\tmin=[10,100,1000]$ Myr.

The number of BNS merger events, $N_i=\{0, 1, \dots\}$, {\it observed} for any given galaxy over a given period of time, $\Delta t$, follows a Poisson distribution based on the merger rate:
\begin{equation}
    P(N_i|\Gamma, t_{\rm min}, \psi_i(z)) = {\rm Poisson}(\dot{n}_i \Delta t) 
    = \frac{(\dot{n}_i \Delta t)^{N_i} e^{-\dot{n}_i \Delta t} }{N_i!}
\label{eqn:poisson}
\end{equation}
We then can simulate $N_i$ directly by drawing it from the Poisson distribution based on the associated rate:
\begin{equation}
    N_i \sim {\rm Poisson}(\dot{n}_i \Delta t)
\end{equation}

Once we have simulated a set of BNS merger events, we determine the constraining power they have on the underlying DTD. Assuming the BNS merger events in each galaxy are independent of each other and the SFHs are known, the corresponding likelihood is
\begin{equation}
    P(\{N_i\}|\Gamma, t_{\rm min}, \{ \psi_i(z) \})
    = \prod_{i=1}^{N_{\rm gal}} P(N_i|\Gamma, t_{\rm min}, \psi_i(z)).
\end{equation}
We further need to marginalize over the uncertainty on the SFH of each galaxy:
\begin{equation}
    P(N_i|\Gamma, t_{\rm min}) = \int P(N_i|\Gamma, t_{\rm min}, \psi_i(z)) P(\psi_i(z)) d[\psi_i(z)].
\end{equation}
We can approximate this integral by averaging over $N_{\rm samp}$ of the samples from the SFH posteriors for each galaxy:
\begin{equation}
    P(N_i|\Gamma, t_{\rm min}) \approx \frac{1}{N_{\rm samp}} \sum_{j=1}^{N_{\rm samp}} \frac{(\dot{n}_{i,j} \Delta t)^{N_i} e^{-\dot{n}_{i,j} \Delta t}}{N_i!}
\end{equation}
The resulting SFH-marginalized posterior of the DTD is therefore given by:
\begin{align}
    P(\Gamma, t_{\rm min} | \{ N_i \}) &\propto P(\{ N_i \} | \Gamma, t_{\rm min} |) P(\Gamma, t_{\rm min}) \\
    &\propto \prod_{i=1}^{N_{\rm gal}} \sum_{j=1}^{N_{\rm samp}} \frac{(\dot{n}_{i,j} \Delta t)^{N_i} e^{-\dot{n}_{i,j} \Delta t}}{N_i!}
\end{align}
where we have assumed that the prior over $\Gamma$ and $t_{\rm min}$ is uniform such that $P(\Gamma, t_{\rm min})$ is a constant.

We are interested in marginalizing over any particular set of events $\{ N_i \}$ associated with the $N_{\rm gal}$ possible host galaxies to forecast possible constraints on the DTD as a function of the \textit{total number} of BNS merger events, $N_{\rm BNS}$. This gives: 
\begin{align}
    &P(\Gamma, t_{\rm min} | N_{\rm BNS}) \nonumber \\
    &= \int P(\Gamma, t_{\rm min} | \{ N_i \}, N_{\rm BNS}) P(\{ N_i \} | N_{\rm BNS}) d(\{ N_i\})
\end{align}
We can approximate this integral using $N_{\rm repeat}$ realizations of the observed BNS merger event counts, conditioned on the total number of events $\sum_i N_i$ being equal to $N_{\rm BNS}$. Combining this with our previous expression then gives:
\begin{align}
    P(\Gamma, t_{\rm min} | N_{\rm BNS}) &\approx \frac{1}{N_{\rm repeat}} \sum_{k=1}^{N_{\rm repeat}} P(\Gamma, t_{\rm min}|\{N_i\}_k) \\
    &\propto \sum_{k=1}^{N_{\rm repeat}} \prod_{i=1}^{N_{\rm gal}} \sum_{j=1}^{N_{\rm samp}} \frac{(\dot{n}_{i,j}\Delta t)^{N_{i,k}} e^{-\dot{n}_{i,j} \Delta t}}{N_{i,k}!}
\end{align}
where again $\sum_{i=1}^{N_{\rm gal}} N_{i,k} = N_{\rm BNS}$ for each realization.

Since the rate per galaxy is generally very small, $\dot{n}_i \ll 1$, we expect the uncertainty to be dominated by variation in the observed counts over the potential $N_{\rm gal}$ host galaxies rather than the uncertainties in each galaxy's SFH. As such, we opt to use the same SFHs used to generate the data to evaluate the posterior rather than trying to marginalize over them directly:
\begin{equation}
    P(\Gamma, t_{\rm min} | N_{\rm BNS}) \sim \sum_{j=1}^{N_{\rm repeat}} \prod_{i=1}^{N_{\rm gal}} \frac{(\dot{n}_{i,j}\Delta t)^{N_{i,j}} e^{-\dot{n}_{i,j} \Delta t}}{N_{i,j}!}
    \label{eqn:posterior1}
\end{equation}

\begin{figure*}
\centering
\includegraphics[width=0.66\columnwidth]{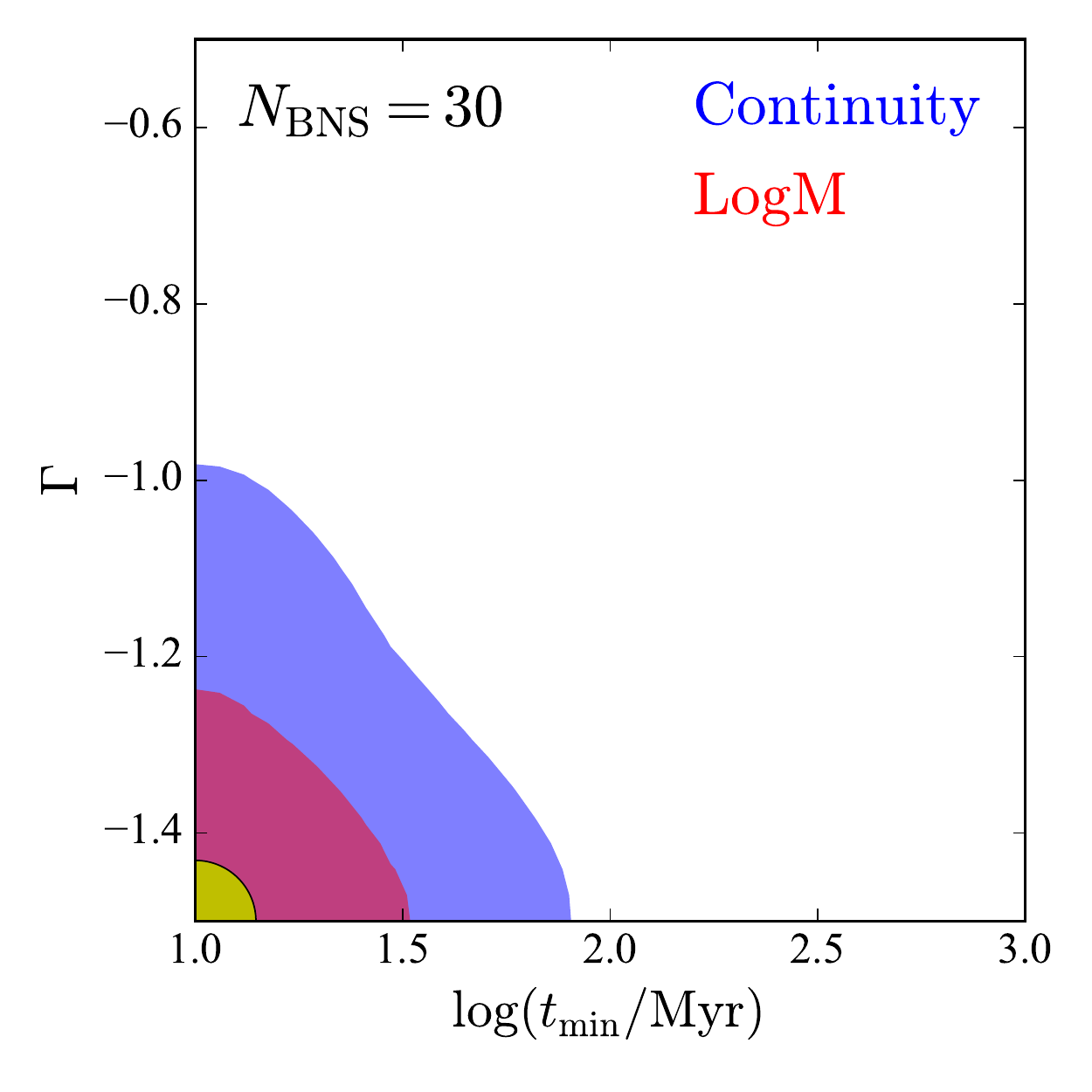}
\includegraphics[width=0.66\columnwidth]{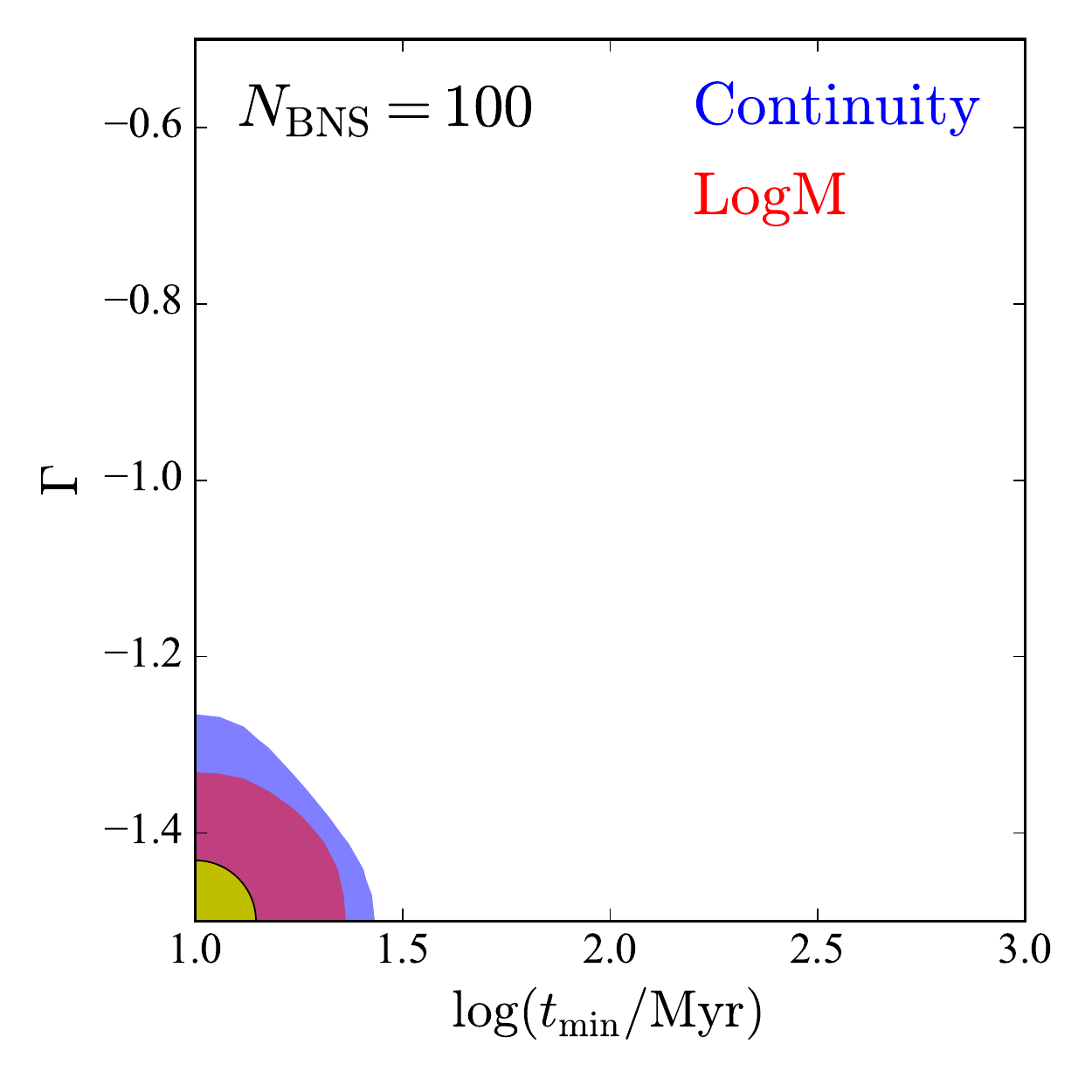}
\includegraphics[width=0.66\columnwidth]{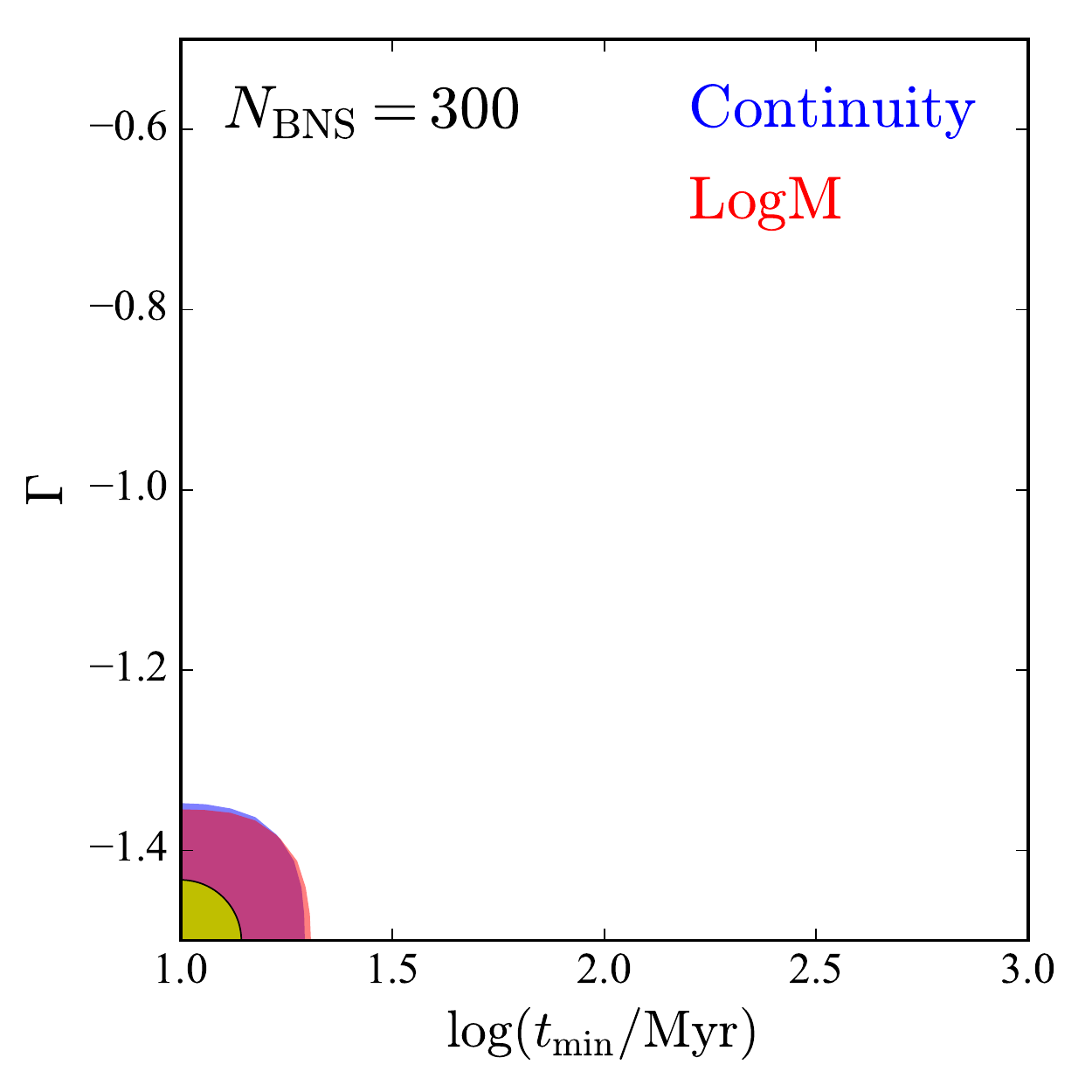}
\includegraphics[width=0.66\columnwidth]{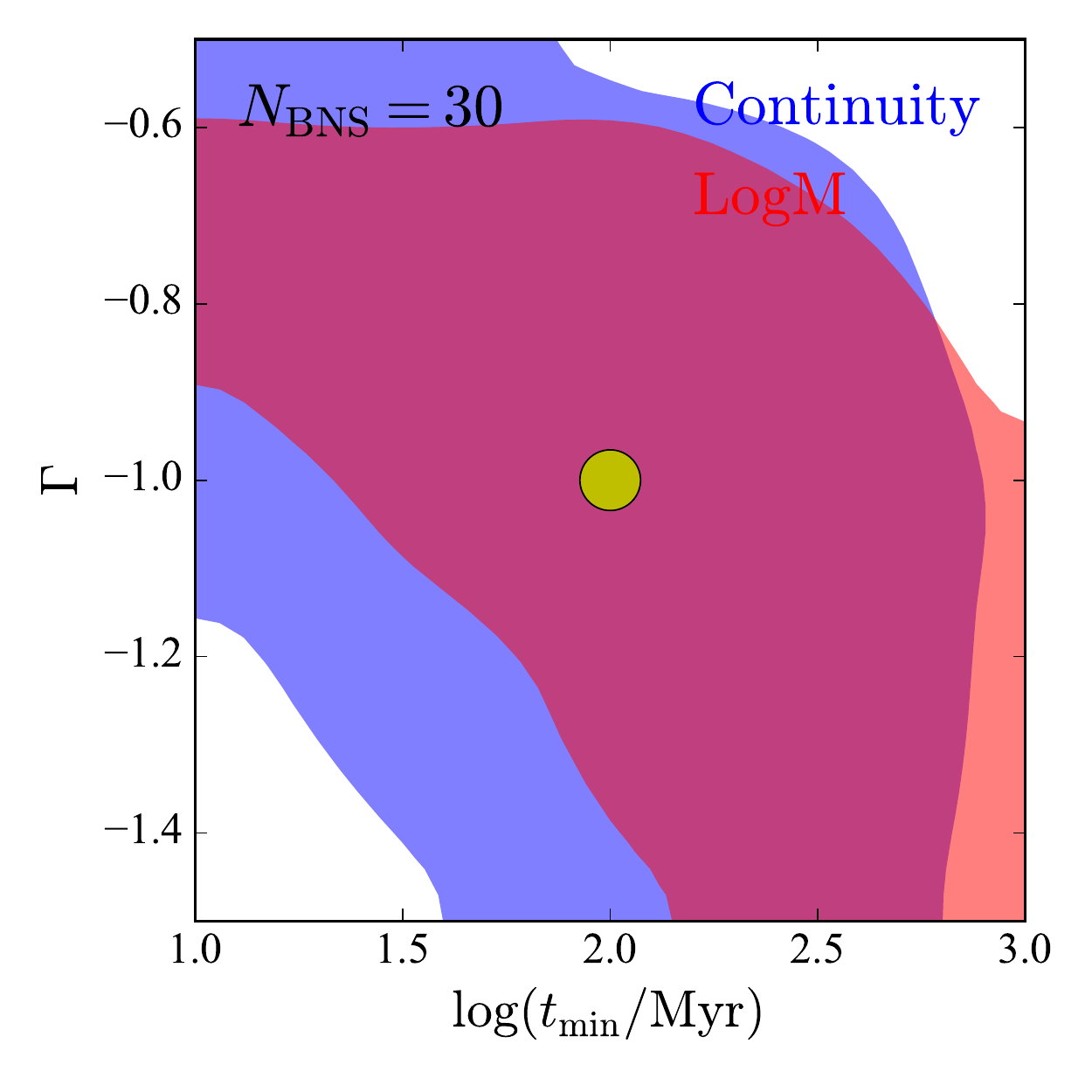}
\includegraphics[width=0.66\columnwidth]{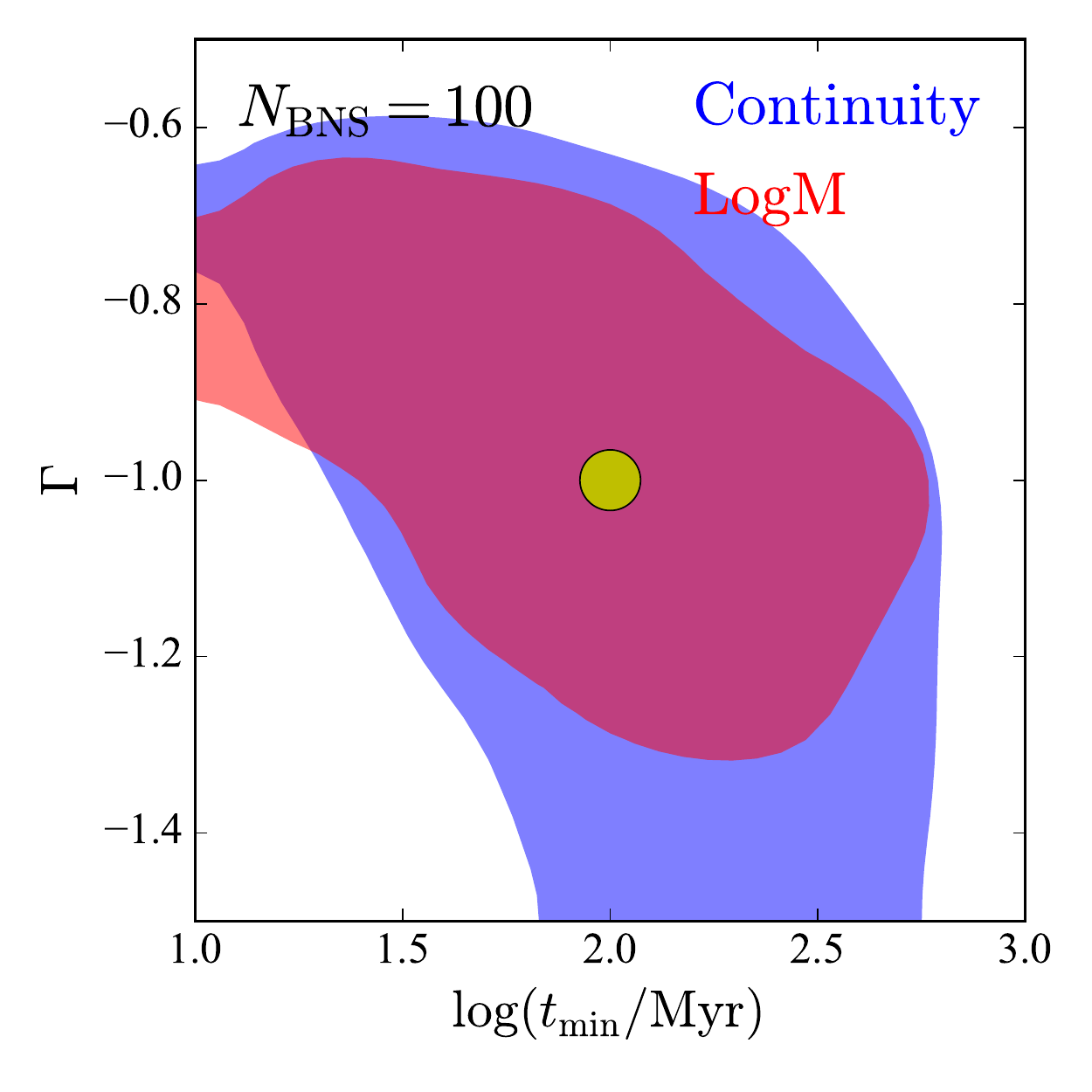}
\includegraphics[width=0.66\columnwidth]{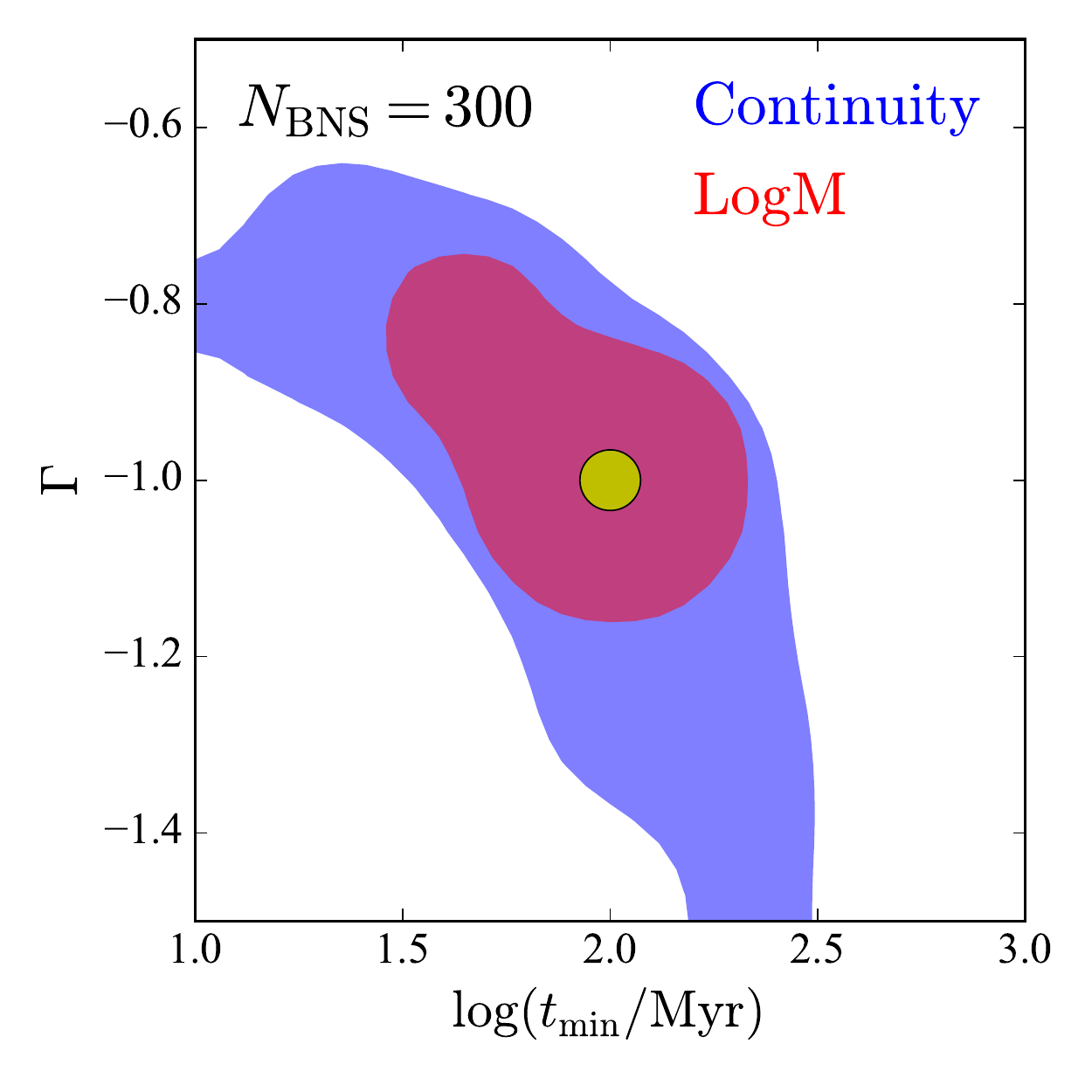}
\includegraphics[width=0.66\columnwidth]{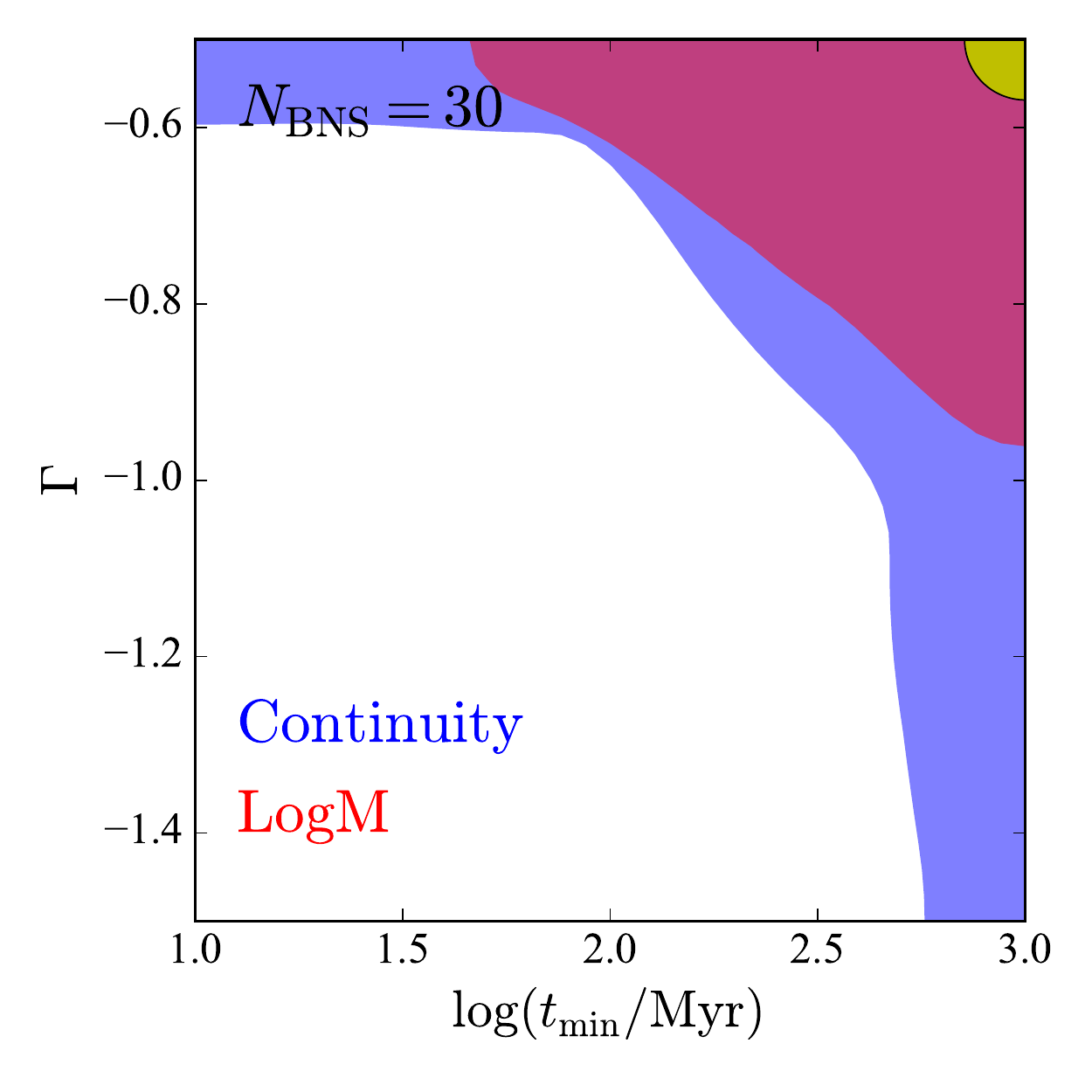}
\includegraphics[width=0.66\columnwidth]{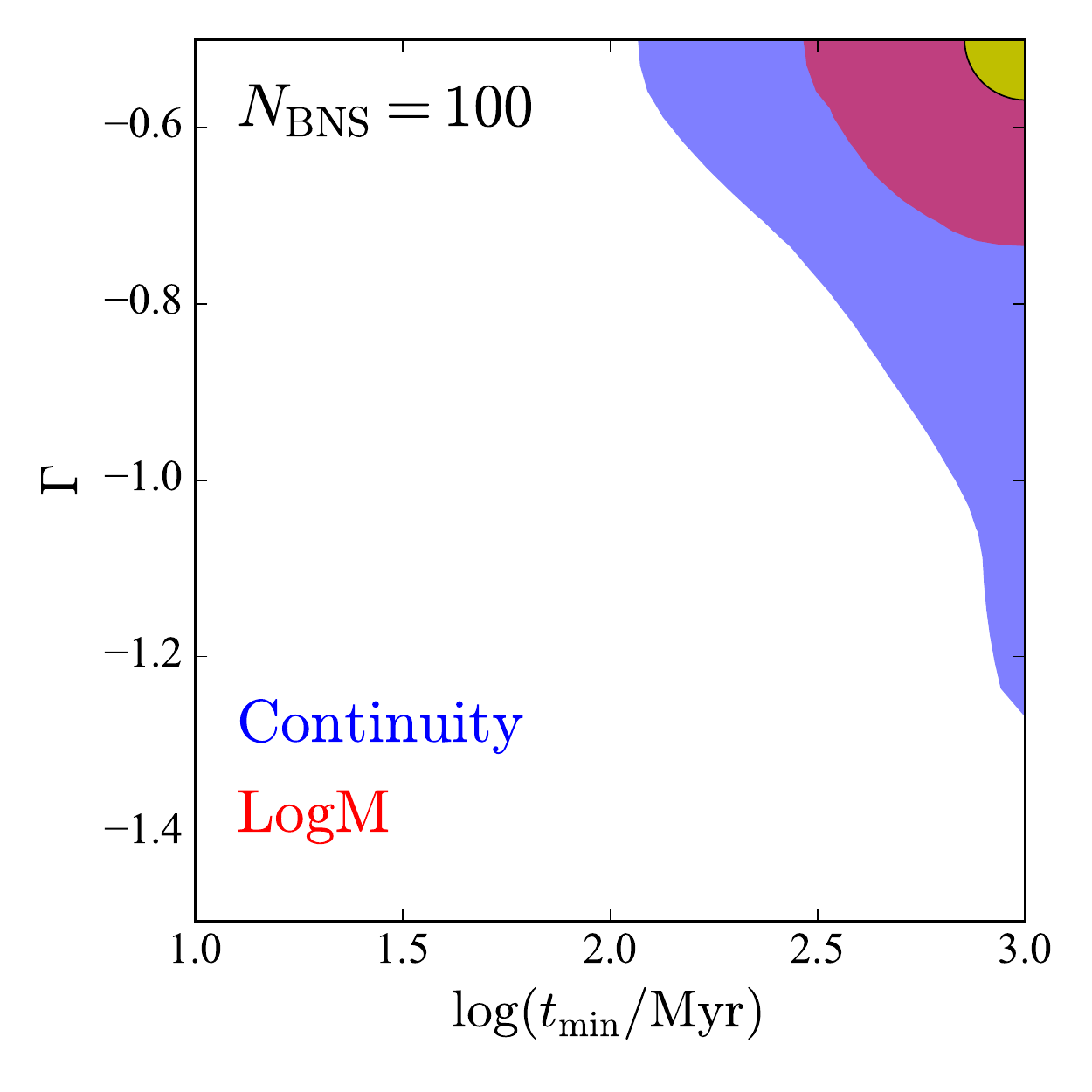}
\includegraphics[width=0.66\columnwidth]{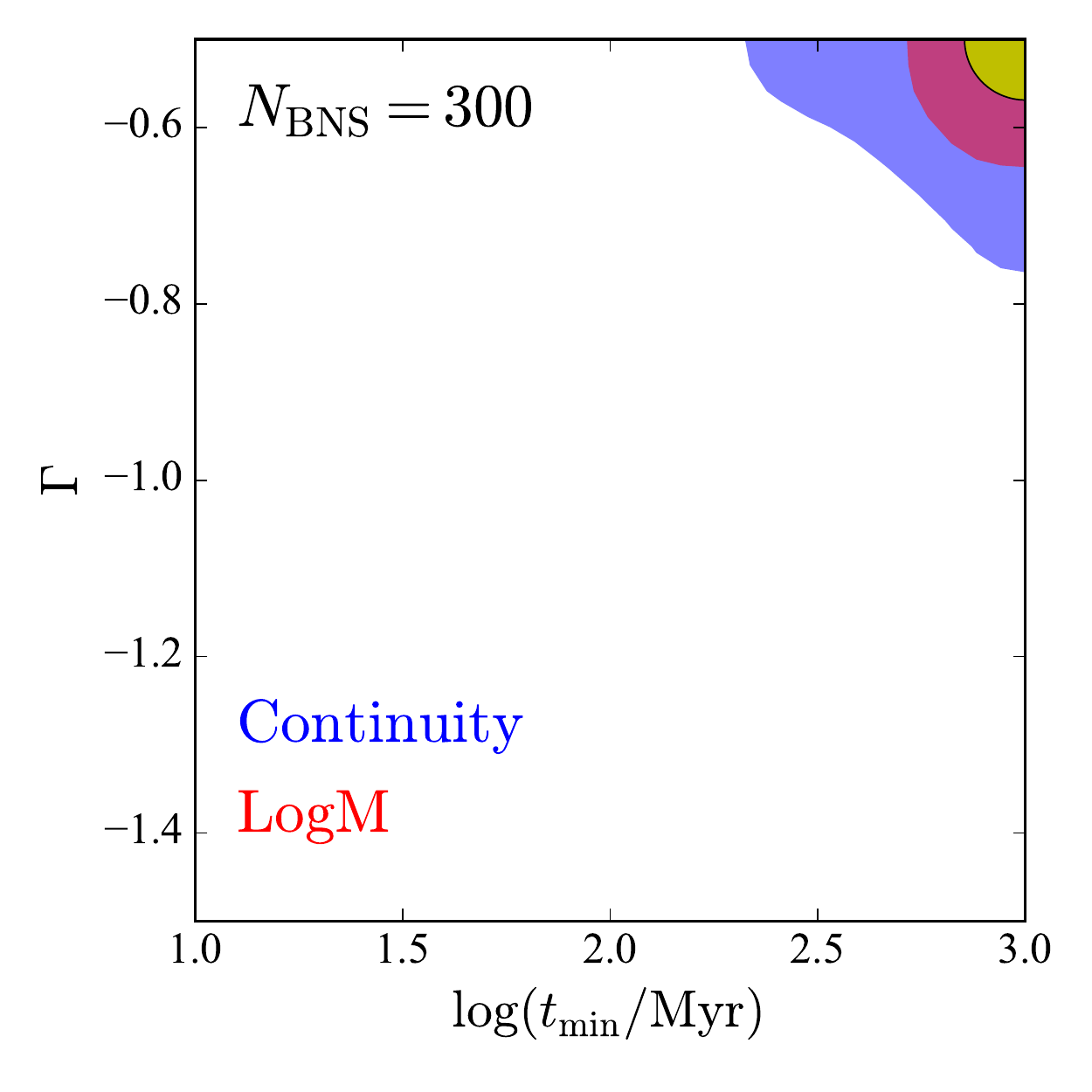}
\caption{The constraint achieved on the parameters of the DTD as a function of the number of host galaxies of BNS merger events ({\it Left}: 30; {\it Middle}: 100; {\it Right}: 300) and for the two choices of SFH priors (red: logM; blue: continuity). In each row the input model is marked with a yellow circle.  {\it Top:} An input DTD with $\Gamma=-3/2$ and $\tmin=10$ Myr. {\it Middle:} An input DTD with $\Gamma=-1$ and $\tmin=100$ Myr. {\it Bottom:} An input DTD with $\Gamma=-1/2$ and $\tmin=1000$ Myr. The contours show the 90\% percentile confidence.}
\label{f:figure_2}
\end{figure*}

It is critical to note that the posterior above is taken over the observed counts of \textit{all} potential galaxy hosts, including those with $N_i = 0$ for which no BNS merger events have been detected. That is because these ``non-detections'' in aggregate contain a non-negligible amount of information in a regime where  $\dot{n}_i\ll 1$ and merger events are rare. To illustrate this, we also compare the ``complete'' posterior distribution $P(\Gamma, t_{\rm min} | N_{\rm BNS})$ derived above with the biased posterior distribution, $\tilde{P}(\Gamma, t_{\rm min} | N_{\rm BNS})$, ignoring the non-detections:
\begin{align}
    \tilde{P}(\Gamma, &t_{\rm min} | N_{\rm BNS}) \sim \sum_{j=1}^{N_{\rm repeat}} \prod_{i=1}^{N_{\rm gal}} \mathcal{I}(N_{i,k} > 0) \frac{(\dot{n}_{i,j} \Delta t)^{N_{i,j}} e^{-\dot{n}_{i,j} \Delta t}}{N_{i,j}!},
    \label{eqn:posterior2}
\end{align}
where $\mathcal{I}(N_{i,j} > 0)$ is the indicator function that evaluates to $1$ if the condition $N_{i,j} > 0$ is true and $0$ otherwise. In general, we expect that ignoring non-detections will bias the inferred DTD, which we discuss in the next section.

We compute the above posterior using $N_{\rm gal} = 1000$ galaxies and $N_{\rm repeat} = 100$ realizations for the 9 different assumed DTDs, and interpolate between their associated $\Gamma$ and $t_{\rm min}$ parameters to obtain the probability for a different pair of $\Gamma$ and $\tmin$ values.

To summarize, our inference procedure is as follows:
\begin{enumerate}
    \item We select a star formation history, $\psi_i(z)$, for every galaxy $i$ from its SFH posterior distribution and compute the corresponding BNS merger rate, $\dot{n}_i$, using Equation~\ref{eqn:ndot}.
    \item We then sample the number of BNS merger events, $N_i$, from the corresponding Poisson distribution based on Equation~\ref{eqn:poisson} for a given timescale $\Delta t$.  We repeat this process until the total number of events is $N_{\rm BNS} = \sum_i N_i$.
    \item We repeat this procedure $N_{\rm repeat}$ times to generate many realizations of the BNS merger events, $\{ N_{i} \}$, for varying SFHs.
    \item We then use the simulated BNS merger events and SFHs from these realizations to compute the DTD posteriors including and excluding galaxies that did not host detected BNS merger events using Equations~\ref{eqn:posterior1} and \ref{eqn:posterior2}, respectively.
\end{enumerate}

\section{Results}
\label{sec:results}

In Figure~\ref{f:priors} we show the BNS merger rate probability distribution function (PDF) of two galaxies from the GAMA survey, with SFHs chosen to peak at early and late cosmic time (right panels), for the 9 different DTD models; the vertical lines in each panel indicate the value of $\dot{n}$ for the mean SFH. We show the SFHs, and merger rate PDFs for both sets of priors (middle panel: continuity; right panel: logM). The SFHs from different assumed priors often do not overlap, illustrating the challenge of inferring SFHs from photometry. In particular, the galaxy in the lower panel illustrates a specific degeneracy where the photometry of star forming galaxies can often be fit equally well with a continuous SFH or with a large burst at $\lesssim 300$ Myr followed by a sharp drop in the star formation rate \citep{leja19}. The figure also clearly indicates how the convolution of the DTDs with different SFHs leads to distinct merger rate PDFs and how the choice of SFH prior affects the resulting merger rate PDFs.

\begin{figure}[!t]
\includegraphics[width=\columnwidth]{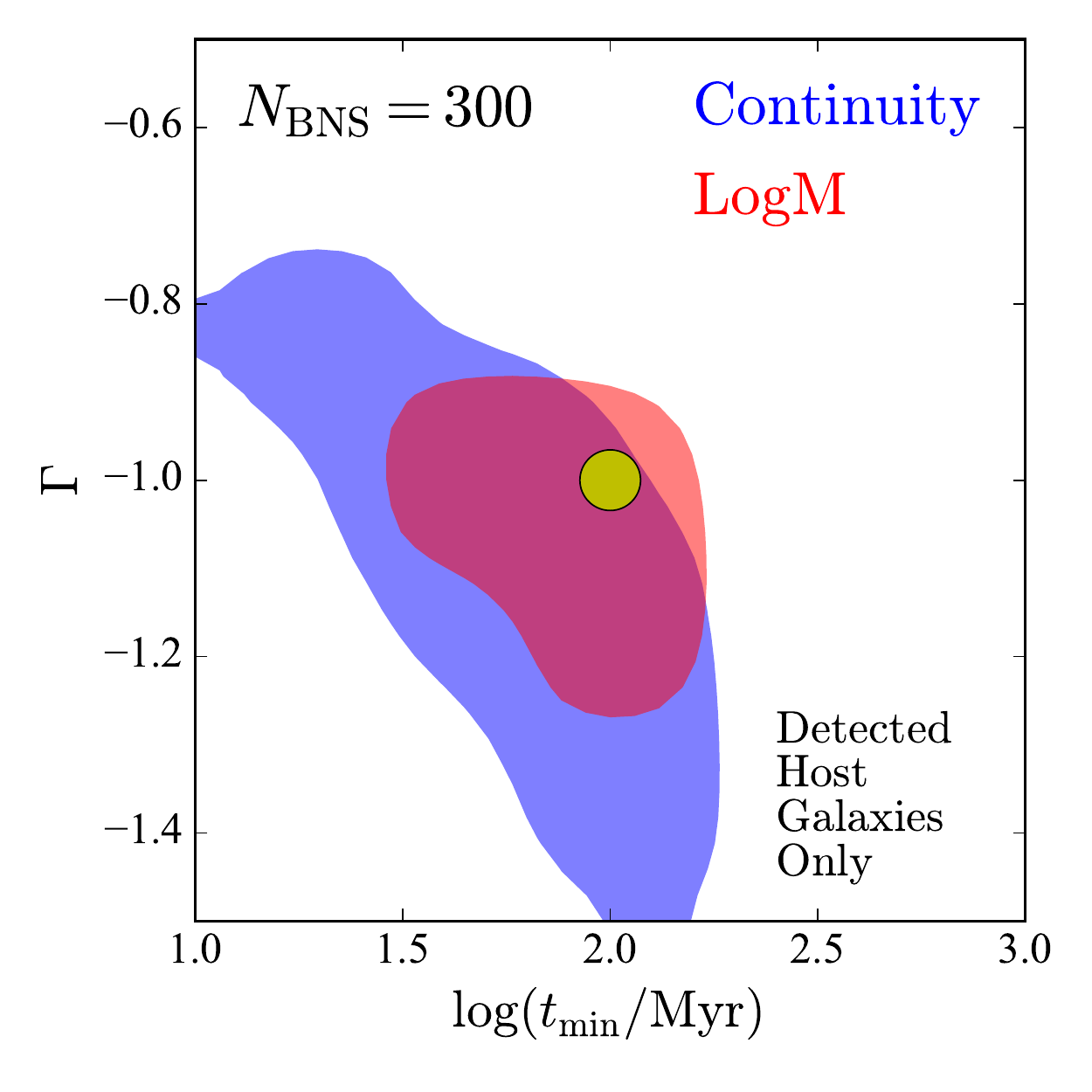}
\caption{Same as Figure~\ref{f:figure_2}, but for just a single injected DTD (yellow circle) and a sample size of 300 host galaxies.  Here we use the SFHs of only the host galaxies and ignore the galaxies that did not host BNS mergers (Equation~\ref{eqn:posterior2}).  We find that the resulting DTD parameters are biased with respect to the input model, but not severely.}
\label{f:figure_3}
\end{figure}

In Figure~\ref{f:figure_2} we show the constraints achieved on the parameters of the DTD model as the number of observed host galaxies increases from 30 to 100 to 300. We show the results for three different injected DTD models and for both the logM and continuity SFH priors. The contours mark the 90\% confidence region. There are several key takeaway points from Figure~\ref{f:figure_2}.  First, in the case that the true DTD prefers short merger timescales, by having a steep power law slope and short $\tmin$ (upper row of Figure \ref{f:figure_2}), the DTD parameters can be reasonably constrained with fewer than $\mathcal{O}(100)$ host galaxies.  Second, in other permutations of the DTD, $\mathcal{O}(300)$ host galaxies may be required to constrain the DTD parameters, but with a lingering degeneracy between $\Gamma$ and $\tmin$.  Third, the logM prior leads to tighter constraints on the DTD parameters compared to the continuity prior because it allows for bursty SFHs that pick more well-defined timescales when convolved with the DTD; the continuity prior smooths the SFH and hence systematically reduces its constraining power.

We note that the results in Figure~\ref{f:figure_2} assume knowledge of the SFHs of all potential host galaxies in the cosmic volume that contains the BNS merger events.  In the case of Advanced LIGO/Virgo at design sensitivity, the luminosity distance range for BNS merger detection is about 200 Mpc, while for the planned A+ and Voyager upgrades this distance is expected to be at least twice as large.  Thus, even at Advanced LIGO/Virgo sensitivity, this requires knowledge of the SFHs of $\sim 10^6$ galaxies, while for A+/Voyager this number increases to $\gtrsim 10^7$ galaxies.  In Figure \ref{f:figure_3} we demonstrate the effect of neglecting the galaxies that did not host BNS mergers, using instead the SFHs of only the actual host galaxies. As expected, the resulting reconstructed DTD parameter distribution is biased with respect to the input model.  However, this bias is not severe, and the resulting degenerate range of $\Gamma$ and $\tmin$ contains the ``true'' answer.  This bias may be acceptable given the need to model the SFHs of only a few hundred galaxies as opposed to $\gtrsim 10^6$ galaxies.

\section{Summary and Conclusions}
\label{sec:conc}

As we have argued in Papers I and II, the DTD of BNS systems can be constrained in two primary ways using GW events: (i) using the properties of BNS merger host galaxies in the local universe, identified via an associated electromagnetic counterpart (Paper I and here); and (ii) using the BNS merger rate as a function of redshift, which requires third-generation GW detectors (Paper II). Here we expand on the method of Paper I, in which we used galaxy scaling relations to relate the mass function of BNS merger host galaxies to the parameters of the DTD.  In particular, we explore the use of the actual SFHs of individual BNS merger host galaxies.

We find that the SFH reconstruction method improves on the use of scaling relations, reducing the required sample size by a factor of $\sim 3-10$, to $\mathcal{O}(100)-\mathcal{O}(300)$.  The exact level of improvement depends on the choice of SFH prior, as well as on the location of the true DTD in the $\Gamma-\tmin$ parameter space.  We further note that accurate reconstruction of the DTD requires knowledge of the SFHs of not only the actual host galaxies but also of the general galaxy population within the relevant cosmic volume ($\sim 3\times 10^7$ Mpc$^3$ for Advanced LIGO and an order of magnitude larger for A+).  However, while using the SFHs of only the host galaxies results in a bias, we find that this bias is not severe.

With the currently allowed range of the BNS merger rate, $110-3840$ Gpc$^{-3}$ yr$^{-1}$ \citep{gwtc1}, it may take a decade or longer to collect a sample of $\mathcal{O}(100)-\mathcal{O}(300)$ host galaxies.  However, there is a reasonable chance that such a sample will be available prior to the advent of third-generation GW detectors.  This will allow for independent determinations of the DTD from the properties of host galaxies in the local universe (Papers I and III) and from the full merger rate redshift distribution (Paper II).

\acknowledgements 
We are thankful to Enrico Ramirez-Ruiz, Evan Scannapieco, and Doug Finkbeiner for helpful discussions. This work was supported by the National Science Foundation under grant AST14-07835 and by NASA under theory grant NNX15AK82G.  The Berger Time-Domain Group at Harvard is supported in part by NSF under grant AST-1714498 and by NASA under grant NNX15AE50G.  J.L. is supported by an NSF Astronomy and Astrophysics Postdoctoral Fellowship under award AST-1701487. MTS is thankful to the Center for Astrophysics | Harvard \& Smithsonian for hospitality, which made this work possible.

\end{document}